\documentclass[5p,times,onecolumn,double-spaced]{elsarticle}

\usepackage[utf8]{inputenc}
\usepackage{graphicx}
\usepackage{dcolumn}
\usepackage{subcaption}
\usepackage{siunitx}
\usepackage{tabularx}
\usepackage{caption}
\usepackage{booktabs}
\usepackage{multirow}
\usepackage{chemformula}
\usepackage{hyperref}
\usepackage{tikz}
    \usetikzlibrary{positioning, shapes, snakes}

\address[thu]{Department of Physics, Tsinghua University, Beijing 100084, China}
\address[hnu]{Institute of Particle and Nuclear Physics, Henan Normal University, Xinxiang 453007, China}
\address[imp]{Institute of Modern Physics, Chinese Academy of Sciences, Lanzhou 730000, China}

\begin{document}

\title{The neutron array of the compact spectrometer  for heavy ion experiments in Fermi energy region}

\author[thu]{Dawei Si} 
\ead{sdw21@mails.tsinghua.edu.cn}

\author[thu]{Sheng Xiao}
\author[thu]{Yuhao Qin}
\author[thu]{Yijie Wang}
\author[thu]{Junhuai Xu}
\author[thu]{Baiting Tian}
\author[thu]{Boyuan Zhang}
\author[thu]{Dong Guo}
\author[thu]{Qin Zhi}
\author[hnu]{Xiaobao Wei}
\author[hnu]{Yibo Hao}
\author[hnu]{Zengxiang Wang}
\author[hnu]{Tianren Zhuo}
\author[imp]{Yuansheng Yang}
\author[imp]{Xianglun Wei}
\author[imp]{Herun Yang}
\author[imp]{Peng Ma}
\author[imp]{Limin Duan}
\author[imp]{Fangfang Duan}
\author[imp]{Junbing Ma}
\author[imp]{Shiwei Xu}
\author[imp]{Zhen Bai}
\author[imp]{Guo Yang}
\author[imp]{Yanyun Yang}

\author[thu]{Zhigang Xiao} 

\ead{xiaozg@tsinghua.edu.cn}

\begin{abstract}
The emission of neutrons from heavy ion reactions is an important observable for studying the asymmetric nuclear equation of state and the reaction dynamics. A 20-unit neutron array has been developed and mounted on the compact spectrometer for heavy ion experiments (CSHINE) to measure the neutron spectra, neutron-neutron and neutron-proton correlation functions. Each unit consists of a $\rm 15\times 15\times 15~cm^3$ plastic scintillator coupled to  a $ \phi=52 ~\rm mm$  photomultiplier. The Geant4 simulation with optical process is performed to investigate the time resolution and the neutron detection efficiency. The inherent time resolution of 212 ps is obtained by cosmic ray coincidence test. The n-$\gamma$ discrimination and
time-of-flight performance are given by $\rm ^{252}Cf$ radioactive source test  and beam test. The neutron energy spectra have been obtained in the angle range $30^\circ \le \theta_{\rm lab} \le 51^\circ$ in the  beam experiment of $^{124}$Sn+$^{124}$Sn at 25 MeV/u with CSHINE. 
\end{abstract}


\maketitle

\section{Introduction}\label{sec.I}
The isospin transport properties carried by neutron from heavy ion collision in the Fermi region are important observations for the study of isospin mechanics and nuclear equation of state (nEOS). The calculation of the transport model shows that the n/p yield ratio in heavy ion collision is sensitive to the density dependence of the symmetry energy\cite{np_ratio1,np_ratio2,np_ratio3,np_ratio4}, and the shape of the neutron and proton correlation function is directly related to the symmetry energy parameters\cite{npcorrelation1,npcorrelation2}, beside, the neutron and proton correlation function can also be used to extract the space-time characteristics of the emission source\cite{space_time1,space_time2}, the nucleus valence neutron density\cite{neutron_distribution}. 

In addition, the free neutron emission can serve as a potential probe to study the short-range correlation effect in nuclei. Various microscopic nuclear many-body approaches\cite{SRC1,SRC2,SRC3,SRC4,SRC5,SRC6,SRC7} indicate that the short-range correlation effect will cause the nucleon momentum distribution in the nucleus to rise above the Fermi surface and generate a high-momentum tail(HMT). Transport model calculations show that HMT will result in a greater average kinetic energy of the nucleon in nuclei as compared to the case of momentum distribution in the
form of free Fermi gas\cite{SRC8}. Therefore, it is necessary to measure the neutron and proton correlation function and energy spectrum simultaneously, which will provide a new way to study isospin transport, symmetry energy and short-range correlation.

In recent years, a compact spectrometer for heavy ion experiment (CSHINE)\cite{CSHINE1,CSHINE2} has been constructed on the Heavy Ion Research Facility in Lanzhou (HIRFL). It consists of an array of silicon strip detector telescopes (SSDTs), three parallel plate avalanche counters (PPACs)  and a CsI hodoscope\cite{SSD,CsI}, which are capable  to detect light charged particles (LCPs), fission fragments (FFs), and high-energy Bremsstrahlung $\gamma$ radiation, respectively. With the high-granularity SSDTs, the angular distribution of the neutron excess of the LCPs and the correlation functions of $Z=1$ isotopes have been measured, revealing the isospin hierarchy of particle emission \cite{CSHINE_Phy1,CSHINE_Phy2, CSHINE_Phy3}. Moreover, the high isotopic identification of the SSDTs enables the observation of the `ping-pong' modality of the particle emissions as an effect of the nuclear symmetry energy \cite{CSHINE_Phy4}. The fission distribution properties has been investigated by the coincidence of the LCPs and the FFs \cite{CSHINE_Phy5}. Using the CsI $\gamma$ hodoscope, the Bremsstrahlung $\gamma$ emission has been observed, giving the signature of the high momentum tail of the nucleons in nuclei \cite{CSHINE_Phy6}.  The missing function so far is the detection of neutrons. To achieve this purpose,  we have designed a neutron array consisting of 20 detection units, and verified its  performance and n-$\gamma$ discrimination ability through cosmic ray, radioactive source test and in-beam experiment. The neutron array makes the CSHINE  capable of detecting neutrons and charged particles simultaneously in heavy ion reactions at Fermi energies.

In this paper, we report the design and test of the neutron array. The detection efficiency is obtained by the simulation based on Geant4. The inherent time resolution of the neutron unit is obtained through the muon coincidence measurement and the ability to distinguish n-$\gamma$ by time-of-flight (TOF) is confirmed by $^{252}$Cf source test. The experiment of $^{124}$Sn+$^{124}$Sn at 25MeV/u has been completed, the preliminary data analysis of the neutron array has been conducted. The paper is organized as following: Section 2 presents the design and simulation of the unit. Section 3 presents the cosmic ray and radioactive source test. Section 4 presents the overall status of beam experiment and the results of neutron array. Section 5 is the summary.

\section{Detectors and simulation}  \label{sec. II}
\subsection{Neutron detectors}\label{sec. II1}
Plastic scintillators have been extensively used in fast neutron measurement\cite{plastic1, plastic2} due to its high neutron detection efficiency and short light decay time. In our neutron array, each neutron unit consists of a $\rm 15~cm\times 15~cm \times 15~cm$ plastic scintillator coupled with a 2-inch photomultiplier tube (PMT) of Hamamatsu R7724, which  is attached to the  socket and base of Hamamatsu E5859-03. For the plastic scintillator, its C-H ratio is 1:1.1, the light decay time is 2.4ns, and the relative light yield is 50$\%$ $\sim$ 60$\%$ of the anthracene crystal. For the PMT, its transit time and transit time dispersion are 29 ns and 1.2 ns, respectively. The surface of the plastic scintillator is covered with polished  aluminum foil of $\rm 15 \mu m$ thickness to reflect photons, and the aluminum foil is packed with black PET plastic to avoid ambient light. The PMT and scintillator are coupled by optical silica gel to improve the light transmission efficiency. Finally, the scintillator and PMT are packaged in $\rm 5 mm$ thick  duralumin shell. Figure 1 (a) presents the schematic plot of one neutron detector unit.

\begin{figure}[!htb]
\includegraphics[width=0.95\hsize]{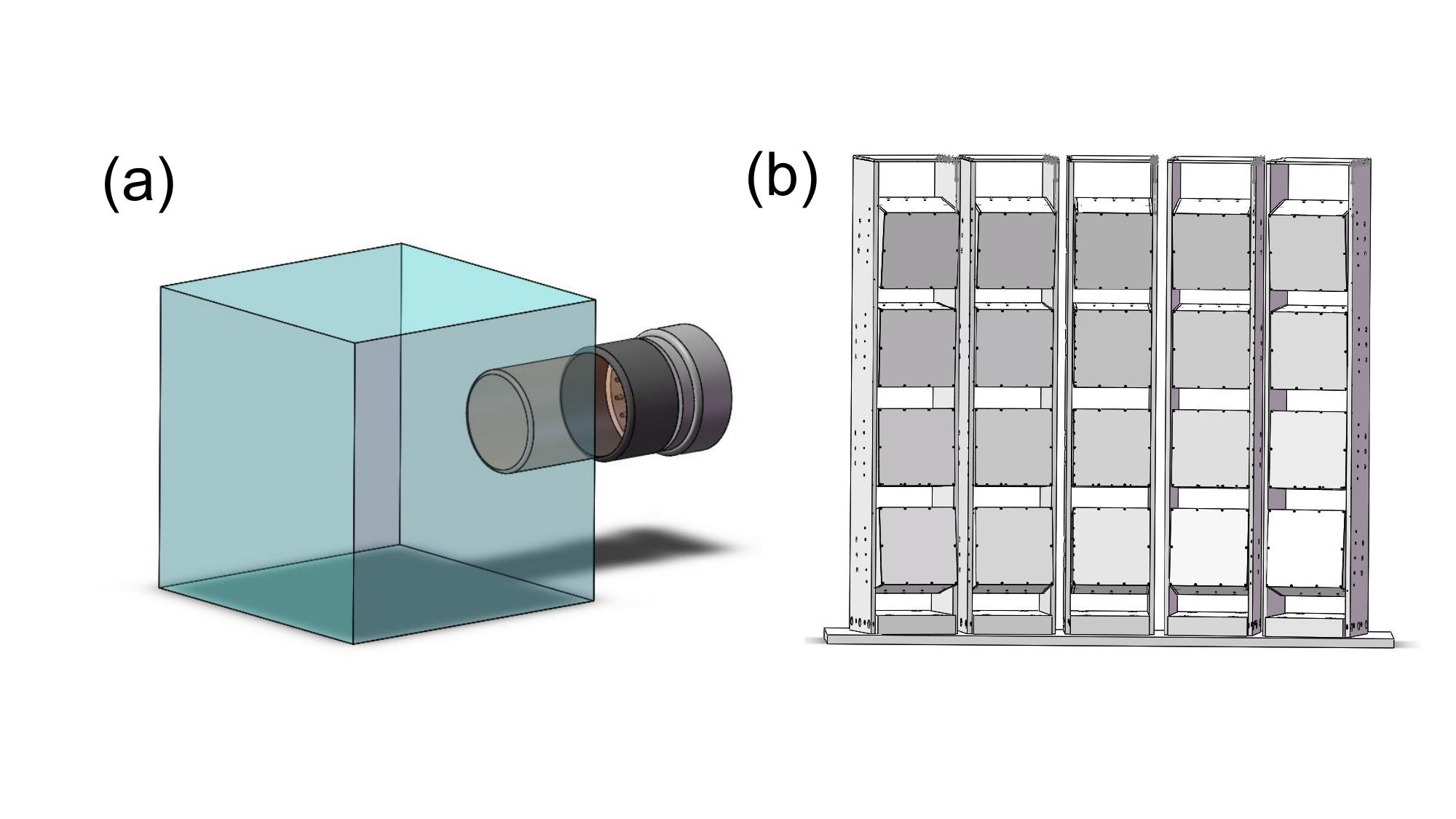}
\caption{(Color online) (a) The structural view of a single neutron detector. (b) the configuration of the neutron array.}
\label{fig1}
\end{figure}

To ensure proper space efficiency and energy resolution in the beam experiment, the 20 neutron units are placed  in 5 columns and 4 rows on the spherical surface being 200 cm  to the target. The units are mounted on a duralumin  frame, and their relative positions are determined by the assembly holes on the frame, as shown in Figure 1 (b). The angular distance between each pair of neighbouring detectors is 6$^{\circ}$  with respect to the target both horizontally and longitudinally.

\subsection{Geant4 simulation}\label{sec. II2}
To estimate the performance of neutron units, a Geant4-based simulation platform is developed. ``FTFP\_BERT\_HP" and ``G4OpticalPhysics" are adopted as the physical process list to describe the interaction of fast neutron in material with high precision neutron model,  and to model the generation and transport of optical photons. Each photon is tracked to its termination, either to be absorbed in propagation or reach on the surface of PMT, where the waveform pulse of given parameters is generated with a certain quantum efficiency. The pulse formed by a single photon is described by\cite{waveform}
\begin{equation}
\label{eq2}
V_{\rm pulse}(t) =
\begin{cases}
G\exp\left[-\frac{1}{2}(\frac{t-t_{\rm i}}{\sigma}+e^{-\frac{t-t_{\rm i}}{\sigma}})\right], & t\leq t_{\rm i}\\
G\exp\left[-\frac{1}{2}\left((\frac{t-t_{\rm i}}{\sigma})^{0.85}+e^{-\frac{t-t_{\rm i}}{\sigma}}\right)\right], & t\textgreater t_{\rm i}\\
\end{cases} 
\end{equation}
where the timing information writes $t_{\rm i}=t_{\rm hit}+t_{\rm trans}$ and $t_{\rm hit}$ represents the time at which a photon hits the PMT. $t_{\rm trans}$ represents the electron transit time of the PMT. The parameter $\sigma$ represents the transit time spread. The final waveform is generated by superimposing all single-photon waveforms, and is recorded with a sampling time interval of 1 ns for digitization. For the boundary characteristics, we used the UNIFIED model\cite{unified} in Geant4 , selected ``dielectric-metal",``polished" option to describe the interface between the plastic scintillator and the  aluminum foil, where the  reflective index is set to 80$\%$\cite{reflectivity}. Fig2 shows the waveform of an incident muon  event in the simulation.
\begin{figure}[!htb]
\includegraphics[width=0.95\hsize]{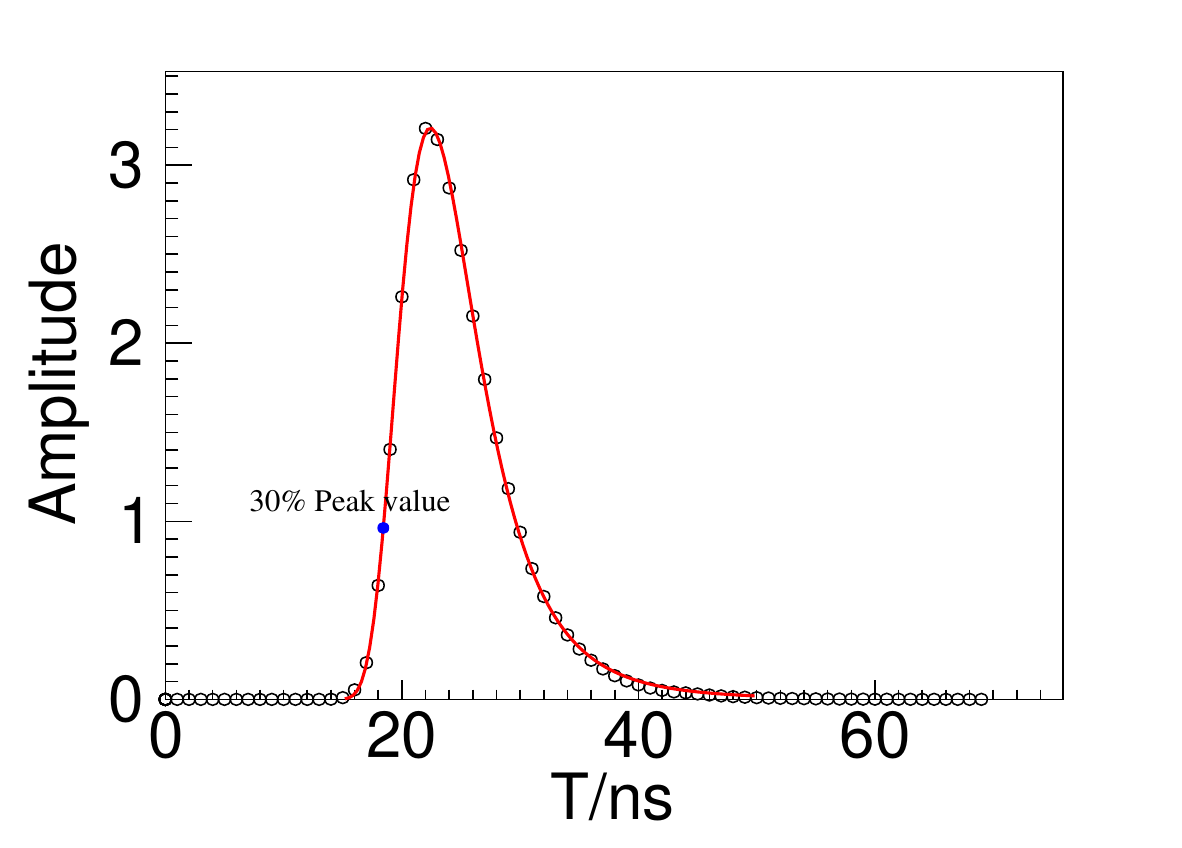}
\caption{(Color online) The waveform of a muon event. the red curve is the result of waveform asymmetric Gaussian fitting, and the time at the $30\%$ peak is selected as the muon arrival time to the unit.}
\label{fig2}
\end{figure}
For time resolution, we have simulated the time difference of the muon passing through two vertically placed neutron units. The vertex of the muon is sampled uniformly in a 30 cm$\times$ 30 cm plane at the top of the upper unit, and its direction and energy is sampled according to the modified Gaisser formula\cite{Gassier}. Figure 3 (a) shows the energy deposition distribution of the muon in the two units. The distribution shows approximately two orthogonal straight bands. The events in the yellow circle indicate the muon crossing the two units vertically, resulting in nearly equal energy deposit. The vertically incident muon events are selected. The signals of the units in coincidence  is then fitted by using the double Gaussian function ( shown in  Figure 2), from which the time corresponding to the 30$\%$ peak height on the rising edge is defined as the arrival time of the muon. Figure 3 (b) presents the spectrum of the time difference of two units fired by vertically incident muons. The Gaussian fit yields  a standard deviation of 249.6(5) ps for the total resolution of the time of flight.

\begin{figure}[!htb]
\includegraphics[width=1.0\hsize]{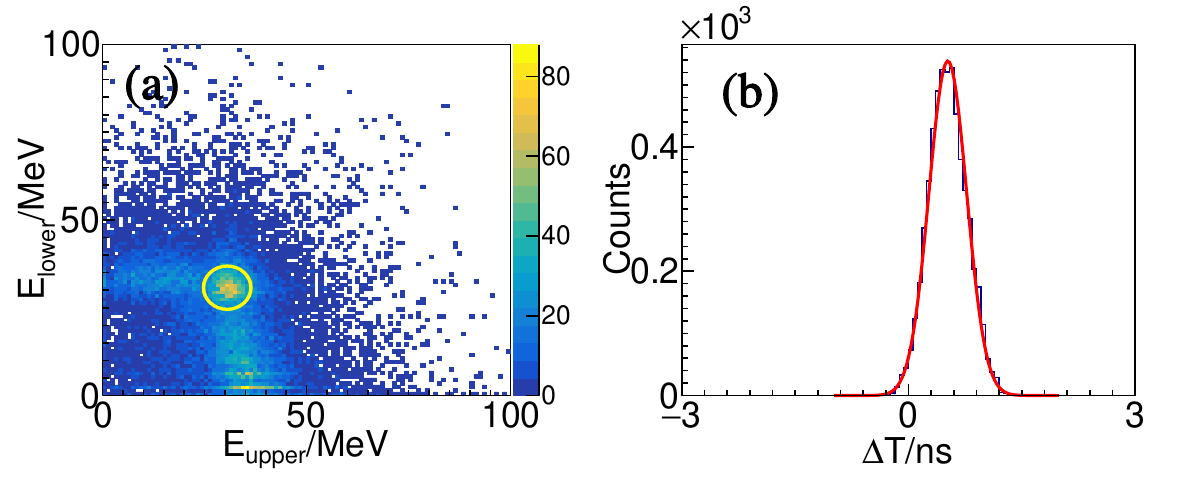}
\caption{(Color online) Geant 4 simulation of the performance of the neutron array to the cosmic muons. (a) Distribution of the energy deposit in the two fired units, (b) the distribution of arriving time difference between the two units for vertically incident muons.}
\label{fig3}
\end{figure}

To estimate the detection efficiency, mono-energetic neutrons are generated isotropically from the target position. The energy is set in the range  from 1 MeV to 100 MeV. The influence of the threshold to the efficiency is also tested in the simulations. The neutron cross-section
data ``G4NDL4.5" is adopted in the simulation. Due to quenching effect, the observed energy($E_{\rm obs}$) is usually lower than the actual neutron deposited energy($E_{\rm real}$), so we use the curve of quenching factor varying with neutron energy of plastic scintillator\cite{quenching} to correct the simulated neutron deposited energy, which is smeared by a Gaussian random dispersion of an energy resolution of 40$\%$, which is typical for plastic scintillators\cite{efficiency1,efficiency2}. The detection efficiency at a certain energy is defined by the ratio of the events where the  energy deposit (mainly induced by the $np$ scattering) exceeds the threshold to the total incident events of a mono energetic neutrons. Figure 4 shows the variation of the detection efficiency as a function of the neutron energy at different threshold. With the increase of neutron energy, the detection efficiency first increases rapidly before it declines as a function of the incident neutron energy. Interestingly,  when the neutron energy is lower than 20 MeV, the detection efficiency exhibits some oscillations due to the capture and fission cross section of low-energy neutrons\cite{efficiency3}. When the incident neutron energy exceeds 20 MeV, the characteristic peak of the energy deposit  spectrum far exceeds the threshold values.

\begin{figure}[!htb]
\includegraphics[width=0.95\hsize]{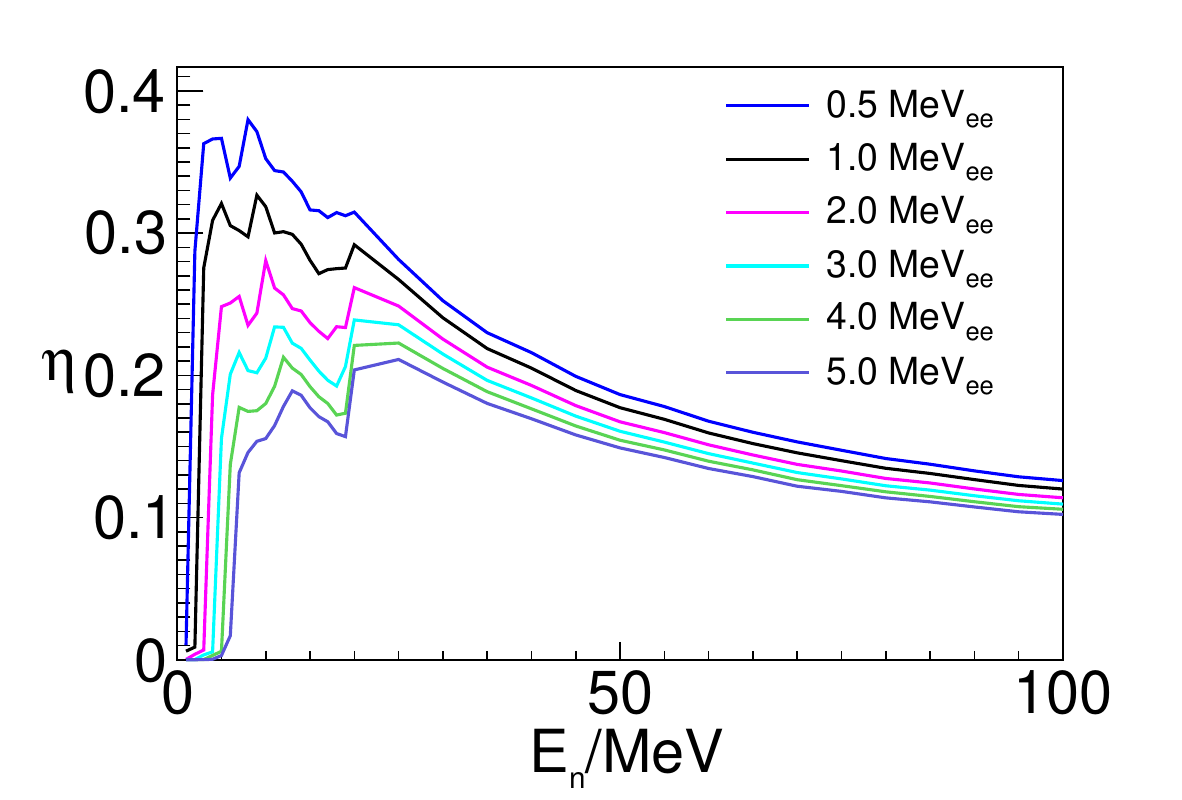}
\caption{(Color online) The detection efficiency as a function of the neutron energy at different threshold.}
\label{fig4}
\end{figure}

\section{Muon and $^{252}$Cf test}\label{sec. III}

Since the cosmic muon is a natural directional background with speed close to the speed of light, the time it takes for a cosmic muon to traverse vertically through two adjacent neutron units is relatively constant, which can be used to verify the time resolution of neutron units by coincidence measurement. We placed the two neutron units vertically to each other, the arriving time of muon was extracted by constant fraction timing discriminator(CFD) and recorded by CAEN TDC V775, and the deposited energy was recorded by CAEN ADC V785 through CAEN spectroscopy amplifier N568E. The deposited energy distribution of upper and lower unit is similar to Figure 3(a), and the time difference distribution was also obtained as shown in Figure 5 by selecting the vertically incident muon events. We fit Figure 5 with the following double Gaussian distribution and get the standard deviation of 301 (41) ps for the time difference.
\begin{equation}
\begin{aligned}
        &C(t) = \frac{N_{1}}{\sqrt{2\pi}\sigma_{1}}\exp(-\frac{(t-\mu_{1})^{2}}{2\sigma_{1}^{2}})&+\frac{N_{2}}{\sqrt{2\pi}\sigma_{2}}\exp(-\frac{(t-\mu_{2})^{2}}{2\sigma_{2}^{2}})\\
        &\sigma = \sqrt{\frac{N_{1}\sigma_{1}^{2}+N_{2}\sigma_{2}^{2}}{N_{1}+N_{2}}}\\
\end{aligned}
\end{equation}
 The inherent time resolution should be 212 ps for single unit, which equal to $\frac{1}{\sqrt{2}}$ of the standard deviation of the time difference distribution.
 \begin{figure}[!htb]
\includegraphics[width=0.95\hsize]{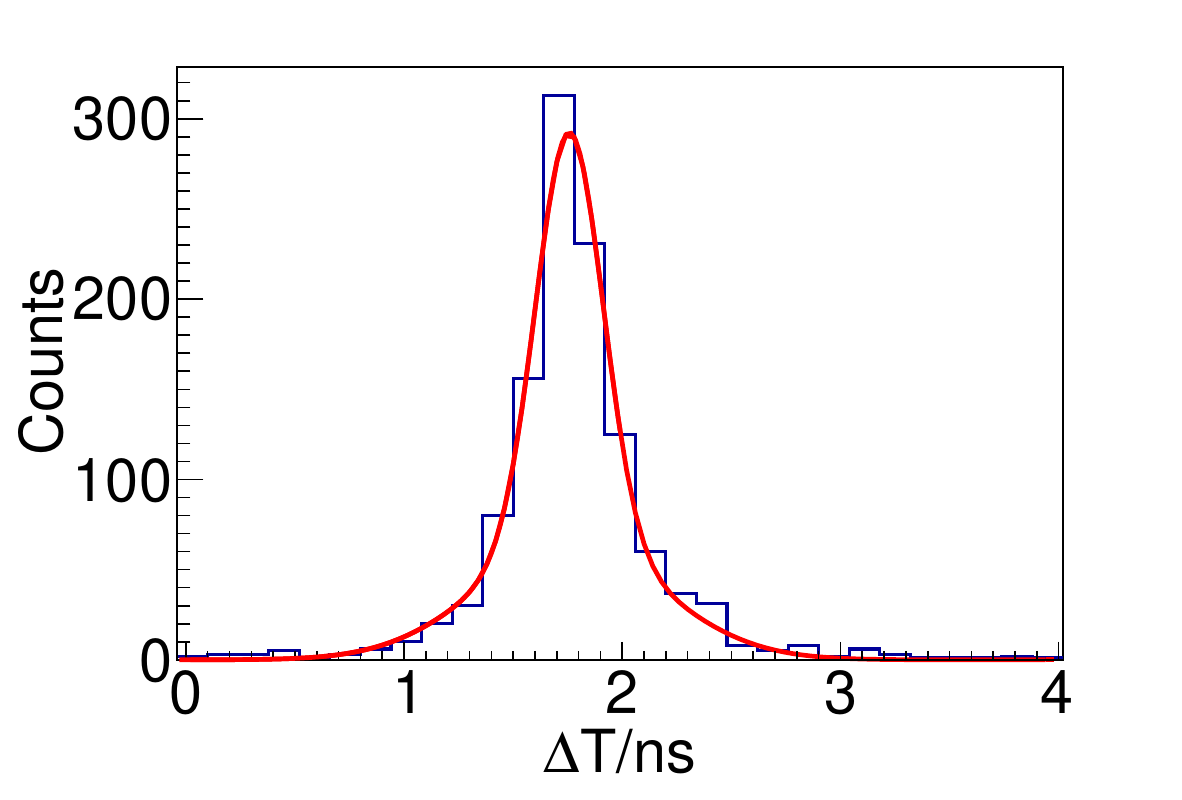}
\caption{(Color online) The time difference distribution of muon coincidence measurement.}
\label{fig5}
\end{figure}

The standard deviation of the  measurement is slightly worse than the simulation result as shown in Fig. 3 (b). The possible reasons are, i) The deviations and fluctuations of the  parameters of the PMTs are at work and ii), there is accidental coincidence background in the  measurement, leading to a long tail on both sides of the time difference distribution, which naturally worsen the timing resolution.
\begin{figure}[!htb]
\includegraphics[width=0.95\hsize]{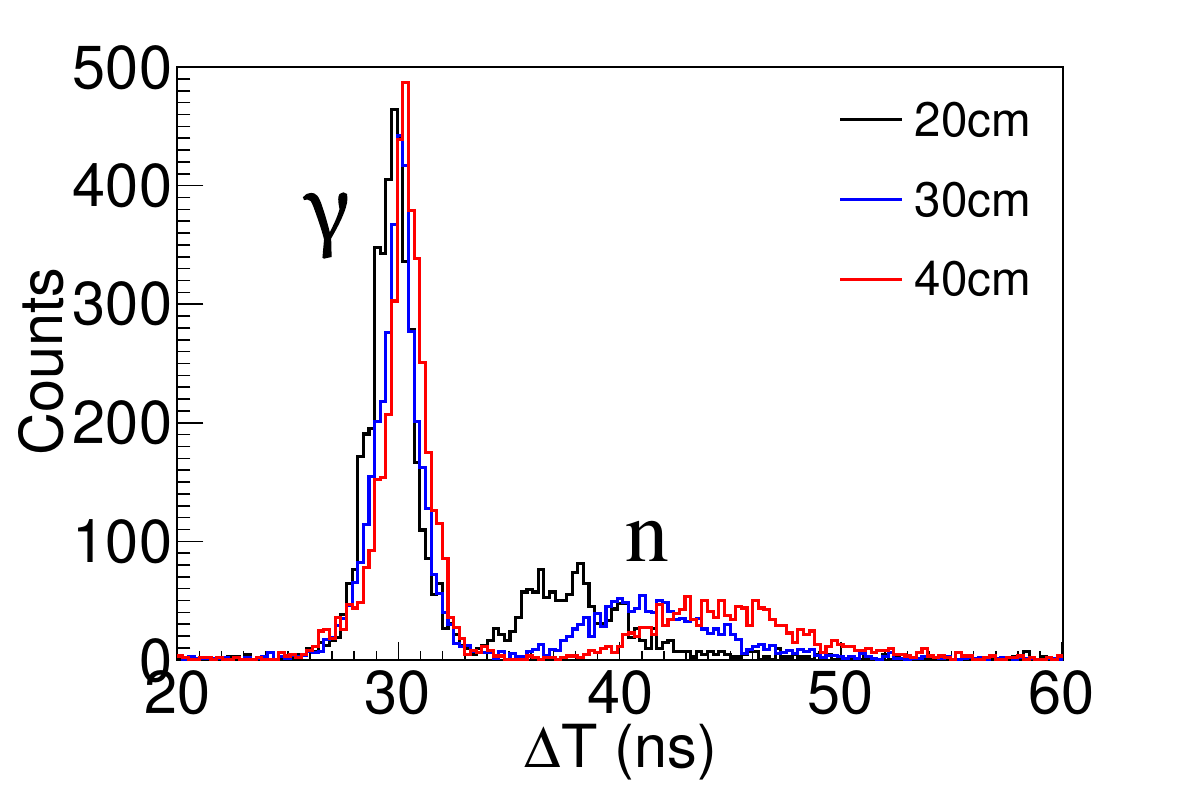}
\caption{(Color online) The TOF spectra for $^{252}$Cf test at different flight distances 20, 30 and 40 cm.}
\label{fig6}
\end{figure}

The ability to distinguish neutron and $\gamma$ ray by time-of-flight has been tested using the $^{252}$Cf fission source which emits  neutrons and $\gamma$ rays simultaneously.  It was performed by setting up a TOF measurement
between a BaF$_{2}$ detector coupled with multi channel plate PMT (MCP-PMT) and the neutron unit. The source was kept adjacent to
to BaF$_{2}$ delivering start timing signal. The surface-to-surface distance between the neutron unit and BaF$_{2}$ varied from 20 cm to 40 cm with a step of 10 cm. The TOF was recorded with data acquisition trigger generated from the coincidence of BaF$_{2}$ and neutron unit. The TOF spectrum obtained at different distance was shown in Figure 6. A clear distinction can be observed in time of flight between neutrons and $\gamma$ rays, as the flight distance increases, the time of flight of neutrons increases more significantly because of its much lower speed compared to $\gamma$ rays. The separation between the two components becomes increasingly pronounced. 

\begin{figure*}[!htb]
\includegraphics[width=0.95\hsize]{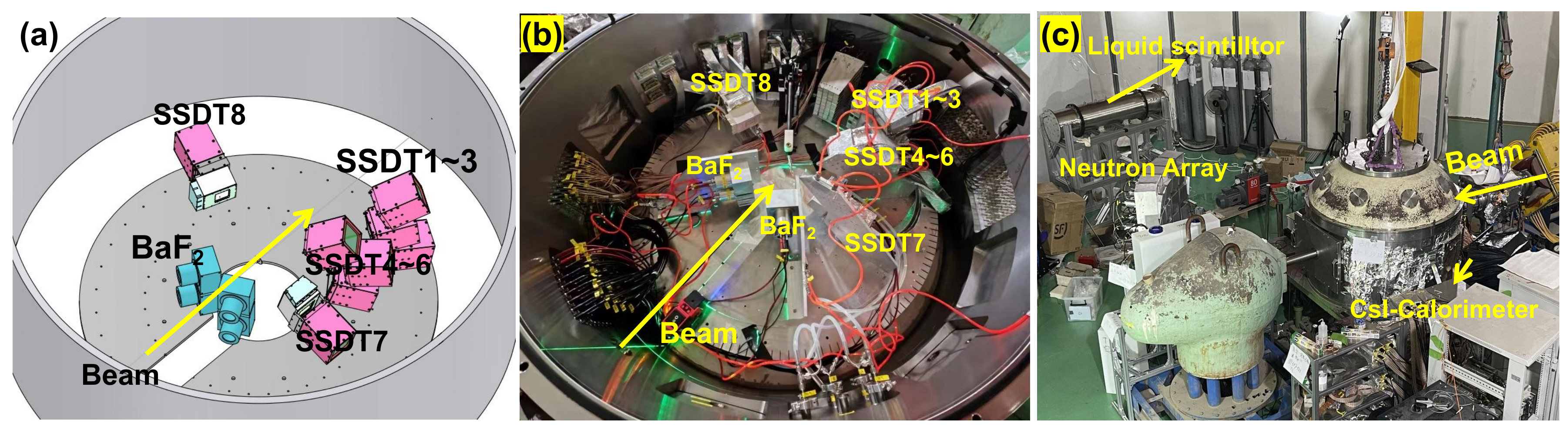}
\caption{(Color online) (a) The schematic diagram of detector setup inside the LSC including Telescope 1$\sim$8 and BaF$_2$ detectors. (b) an image of detector setup inside the LSC. (c) an image of detector setup outside the LSC including neutron array, CsI-calorimeter and liquid scintillator.}
\label{fig7}
\end{figure*}

\section{Beam experiment and Performance}  \label{sec. III}
Finally, we present the performance of the neutron array in the experiment of $^{124}$Sn+$^{124}$Sn at 25 MeV/u, which was performed at the final focal plane of the radioactive ion beam line at Lanzhou (RIBLL1) in 2024. The $^{124}$Sn beam was delivered by the cyclotron of the heavy ion research facility at Lanzhou (HIRFL), bombarding on a $^{124}$Sn
target with the thickness of 1 mg/cm$^{2}$. The detector setup is described below.

\subsection{Detector Setup}  \label{sec. III1}
In the experiment, eight light charged particle telescopes and four BaF$_{2}$ detectors were installed in the Large Scattering Chamber (LSC) with 1.5 m inner diameter located at the last focal plane of RIBLL. Figure 7 presents a picture of the detector setup in the experiment. Telescope 1$\sim$6 are mounted to measure  the charged particles of  $Z\leq 6$  in the polar angular range $25^\circ < \theta_{\rm lab} < 65^\circ$ and azimuthal angular range $40^\circ < \phi_{\rm lab} < 120 ^\circ$ . Each telescope consists of a single-sided silicon strip detector (SSSD), a double-sided silicon strip detector (DSSD) and a $3\times 3$ CsI(Tl) scintillator array. For the performance of the telescopes, one can refer to \cite{SSD} . Telescope 7$\sim$8 consist of ionization Chamber, DSSD and a $3\times 3$ CsI array \cite{chamber}, mounted at $\theta_{\rm lab} =90^\circ$ and $-70^\circ$, respectively, to measure the evaporated charged particles. A 15-unit CsI hodoscope, namely CSHINE-GAMMA \cite{CsI}, surrounded by thick plastic scintillators, was mounted out of the LSC at $\theta_{\rm Lab}$ = -110$^\circ$ and 115 cm to the target, to measure the Bremsstrahlung $\gamma$-rays produced in the heavy ion reactions.   

In addition to the existed detectors, The $4 \times 5$ neutron array and one  liquid scintillator barrel have been newly mounted on CSHINE  to measure the neutrons. The liquid scintillator was placed at $\theta_{\rm lab} = 80^\circ$ and  513 cm away from the target. The geometrical parameters ($R$, $\theta_{\rm lab}$, $\phi_{\rm lab}$)  of the $\rm BaF_2$ detectors and the neutron array are listed in Table 1 and 2 respectively. Here  $R \rm(cm)$ is the distance from the center of scintillator to the target,  $\theta_{\rm lab}$($^\circ$) and $\phi_{\rm lab}$($^\circ$) represent the polar angle and the azimuthal angle in laboratory, respectively. 

\begin{table*}[htbp]
\centering
\caption{The geometrical parameters ($R$, $\theta_{\rm lab}$, $\phi_{\rm lab}$) for BaF$_2$ $\rm T_{0}$ detectors}
\begin{tabular}{cccc}
\toprule
BaF$_2$-1 & BaF$_2$-2 & BaF$_2$-3 & BaF$_2$-4\\
\midrule
(18, 145, -57) & (18, 145, -112) & (18, 145, 57) & (18, 145, -112)\\
\bottomrule
\end{tabular}
\end{table*}

\begin{table*}[htbp]
\centering
\caption{The geometrical parameters ($R$, $\theta_{\rm lab}$, $\phi_{\rm lab}$) for each unit of neutron array }
\begin{tabular}{c|ccccc}
\toprule
     &  column-1 & column-2 & column-3 & column-4 & column-5 \\
\hline
raw-1 & (220.5, 30, 104) & (220.7, 35, 102) & (220.7, 41, 101) & (220.7, 46, 100) & (220.5, 52, 99) \\
\hline
raw-2 & (221.1, 29, 93) & (221.4, 34, 93) & (221.4, 40, 92) & (221.4, 46, 92) & (221.2, 51, 92)\\
\hline
raw-3 & (221.7, 29, 81) & (221.9, 35, 83) & (222.0, 40, 83) & (221.9, 46, 84) & (221.7, 51, 85)\\
\hline
raw-4 & (222.1, 30, 70) & (222.3, 36, 73) & (222.4, 41, 75) & (222.3, 46, 76) & (222.1, 52, 77)\\
\bottomrule
\end{tabular}

\end{table*}

\begin{figure}[!htb]
\centering
\includegraphics[width=0.9\hsize]{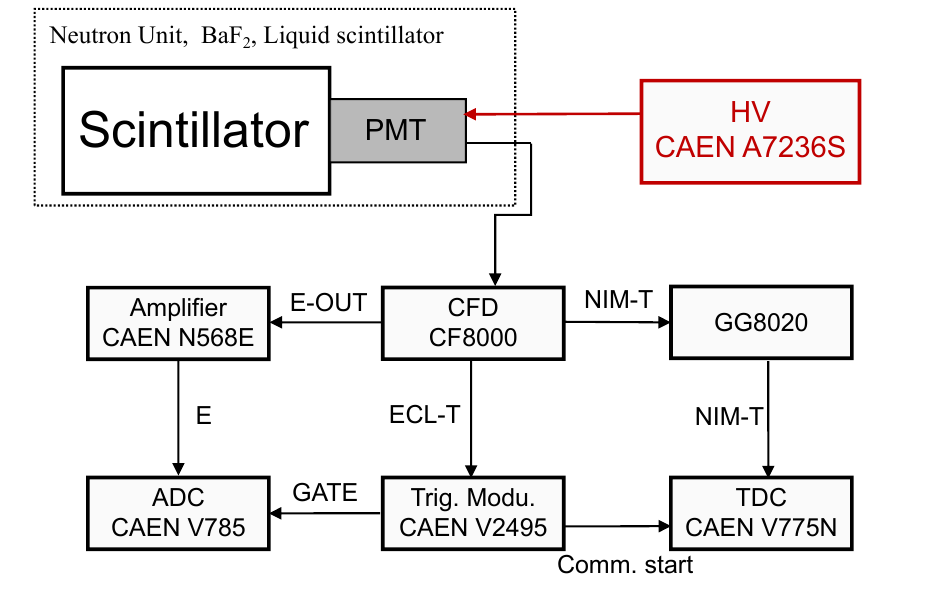}
\caption{(Color online) Readout electronic system of Neutron array, BaF$_{2}$ and liquid scintillator detector}
\label{fig8}
\end{figure}
 
\begin{figure*}[!htb]
\centering
\includegraphics[width=0.9\hsize]{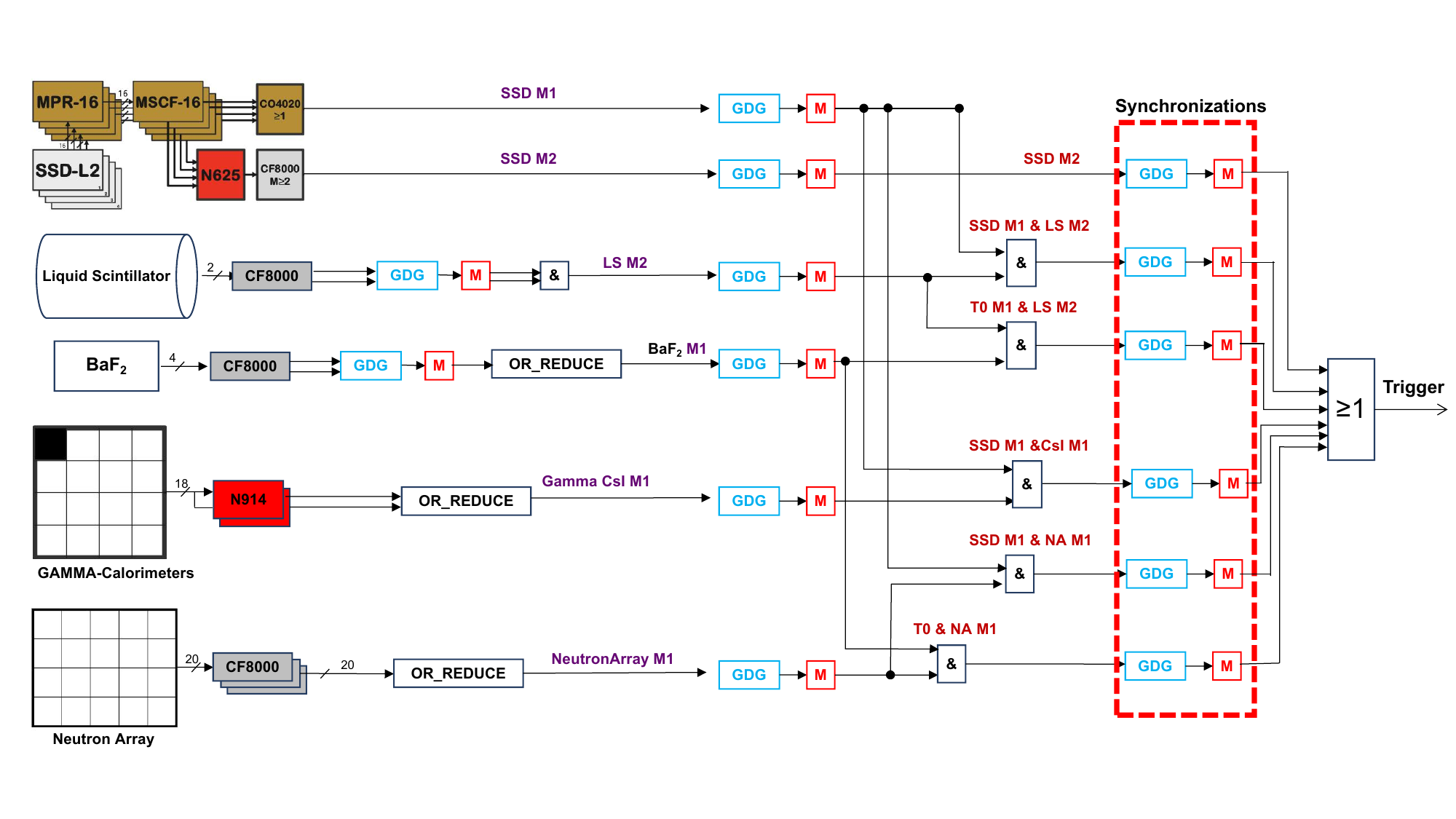}
\caption{(Color online) Logic schematic diagram of beam experiment trigger system}
\label{fig9}
\end{figure*}

The readout electronic system of the neutron array, the $\rm BaF_{2}$ $T_0$ detectors and the liquid scintillator detector in the experiment is presented in Figure 8. The signal from each PMT is transferred to a constant fraction   discriminator (CFD)  and the thresholds of the CFDs for neutron array were set to a common level of $-60$ mV, corresponding to the $\gamma$ energy around 1 MeV in the beam experiment. The NIM logic fast timing output was transferred to the  gate and delay generator (GDG) from the front panel of CFD to set a suitable delay before be digitized by the time-digit converter (TDC) CAEN V775N. And the ECL logic fast timing output was transferred to the trigger unit of CAEN V2495 from the back panel of the CFD to generate a global experiment trigger based on
FPGA technology. The negative-polarized amplitude signals from the back panel of CFD  which is a copy of the input PMT signal, were further transferred to CAEN N568E spectrometer amplifier, the OUT signals of N568E were digitized by the amplitude-digit converter (ADC) CAEN V785. Both the time and amplitude information were saved by the data acquisition (DAQ) system. For the electronics of the telescopes and CsI hodoscope, one can refer to\cite{CSHINE1, CsI}.

To accomplish the physical goals mentioned in Section 1, a global trigger has been generated by the trigger module V2495 as shown in Figure 9.  All the sub-detectors produced logic signals by front-end electronics including:(1) NA.M1: the one-body neutron signal provided by neutron array. (2) T0: the logic signal with at least one $\rm BaF_2$ detector fired. (3)SSD.M1 and SSD.M2: The one-body and two-body light charged particle signal provided by the DSSD of the  telescopes. (4) CsI.M1: the one-body $\gamma$ signal provided by CsI-calorimeter and (5) LS.M2: the coincidence signal of the PMTs at both ends of liquid scintillator detector.
The above five kinds of discrimination signals 
were used to produce six coincidence logical signals as shown in Figure 9, and the global trigger signal is obtained by the OR operation of the six logical signals. The trigger system has been implemented based on the Field Programmable Gate Array (FPGA) technology. For the details of the FPGA based trigger system, one can refer to\cite{CSHINE2}. The number of events acquired with different trigger is shown in Table 3.
\begin{table}[htbp]
\centering
\caption{The event numbers for different trigger condition }
\begin{tabular}{c|c}
\toprule
trigger condition & number of events\\
\hline
$\rm  T_0$ \& NA.M1 & 4300320\\
\hline
SSD.M1 \& NA.M1 & 75229844\\
\hline
$\rm  T_0$ \& LS.M2 & 87193\\
\hline
SSD.M1 \& LS.M2 & 852076\\
\hline
SSD.M1 \& CsI.M1 & 20534609\\
\hline
SSD.M2 & 75229844\\
\bottomrule
\end{tabular}
\end{table}

\subsection{Neutron Spectrum}  \label{sec. III2}

We concentrate on the analysis of the neutron array data here. To understand the performance of the neutron array in the experiment, around 4 million neutron events were analysed by selecting ``$\rm T_0 \& NA.M1$" trigger condition. Due to the manual adjustment uncertainty of the GDG, there is small difference in delay time between different $\rm BaF_2$ and neutron units. Formula (2) was used to fit the $\gamma$ peak on the TOF spectrum of all the combination of BaF$_2$ and neutron units, Figure 10 (a) shows the TOF spectrum of the $\rm 5^{th}$ neutron unit as an example by shifting all the $\gamma$ TOF peaks to zero.   The abscissa is ${\rm TOF}-L/c$, where $L$ is the distance between the unit and the target position and $c$ is the speed of light. It is shown that the two components of $\gamma$ rays and neutrons are separated clearly. The overall experimental  time resolution $\sigma_{\rm exp}=744\pm40$  ps was obtained by fitting the $\gamma$ peak in Figure 10 (a) with formula (2), which is worse than the cosmic ray test result for a single unit, because in the cosmic ray test, the muon  passes vertically through two units and the time variation caused by the variation of the incident position is relatively constant for the two fired units. However the fire position by the $\gamma$ rays originating from the reaction  is varied due to the size of the unit, which is 15 cm in each dimension corresponding to about 700 ps uncertainty. In addition, In Figure 10 (a), an abnormal peak appears in the time range between 4 and 14 ns forms an insignificant, which originates seemingly from the background produced by a tiny portion of the beam particles hitting the vacuum pipe due to the long free space from the last quadrupole doublet to the target.

The neutron energy spectrum is obtained from the TOF distribution by subtracting the contributions of the $\gamma$ rays, the backgrounds and the small abnormal peak. 
For the high energy part with the TOF ranging in $-2.5 \sim 16$  ns, we assume that the TOF distribution follows an  exponential distribution\cite{Source1, Source2, Source3} and use  an asymmetric Gaussian distribution to fit the abnormal peak. The $\gamma$ ray peak was fit by  a double Gaussian distribution. For the low energy part with TOF in  60 ns $\sim$ 150 ns, the exponential distribution with the superposition of a uniform background is adopted to describe the  low energy neutrons. The ansatz (3) is used to describe the boundaries of the TOF distribution  and to remove the abnormal peak and the background in the coincidence window. 

\begin{eqnarray}
\hspace{-1cm}
\begin{aligned}
f(t)&= p_{\rm n_0}\exp(p_{\rm n_1}t)+p_{\gamma_0}\exp\left[-\frac{(t-p_{\gamma_1})^{2}}{2p_{\gamma_2}^{2}}\right]+p_{\gamma_3}\exp\left[-\frac{(t-p_{\gamma_1})^{2}}{2p_{\gamma_4}^{2}}\right] \\
&+p_{\rm ab_0}\exp\left[-\frac{(t-p_{\rm ab_1})^2}{\left(p_{\rm ab_2}+p_{\rm ab_3}(t-p_{\rm ab_1})\right)^2}\right]\\
&+p_{\rm bkg},\quad t \in [{\rm -2.5 ns,16 ns}]\\
f(t)&= p_{\rm n_2}\exp(p_{\rm n_3}t)+p_{\rm bkg},\quad t \in [{\rm 60 ns,150 ns}]    \\
\end{aligned}  
\end{eqnarray}

Totally there are 14 fitting parameters in ansatz (3).  The parameters $p_{\rm n_0} \sim p_{\rm n_3}$ are introduced for the description of  the neutron spectrum, $p_{\gamma_0} \sim p_{\gamma_4}$ are for the $\gamma$ peak,  $p_{\rm ab_0} \sim p_{\rm ab_3}$ are for the tiny abnormal peak and  $p_{\rm bkg}$ is used to describe the uniform background on the TOF spectrum, respectively.

The fitting results are shown in Figure 10 (a) for one single unit. It is shown that  the high and low energy boundaries of the TOF spectrum are well reproduced.

Figure 10 (b) presents the energy spectra of different components.  By subtracting  the abnormal peak (blue solid) and the background (black dashed)  components, one can get the clean neutron energy spectrum (pink squares). It is shown that neither the background nor the tiny abnormal peak brings significant contribution to the neutron spectrum. The three arrows pointing to   100, 200, 300 MeV,  corresponding to the TOF  of 17.2, 12.9, 11.3 ns, respectively. Considering the time  uncertainty of about 1 ns, the energy uncertainties  at the three energy points are derived as  15, 50 and 106 MeV, respectively. With these uncertainties, the very high energy tail of the spectrum is problematic, yet the spectrum below 100 MeV is reliable and the influence of the  abnormal peak is insignificant.

\begin{figure}[!htb]
\includegraphics[width=0.95\hsize]{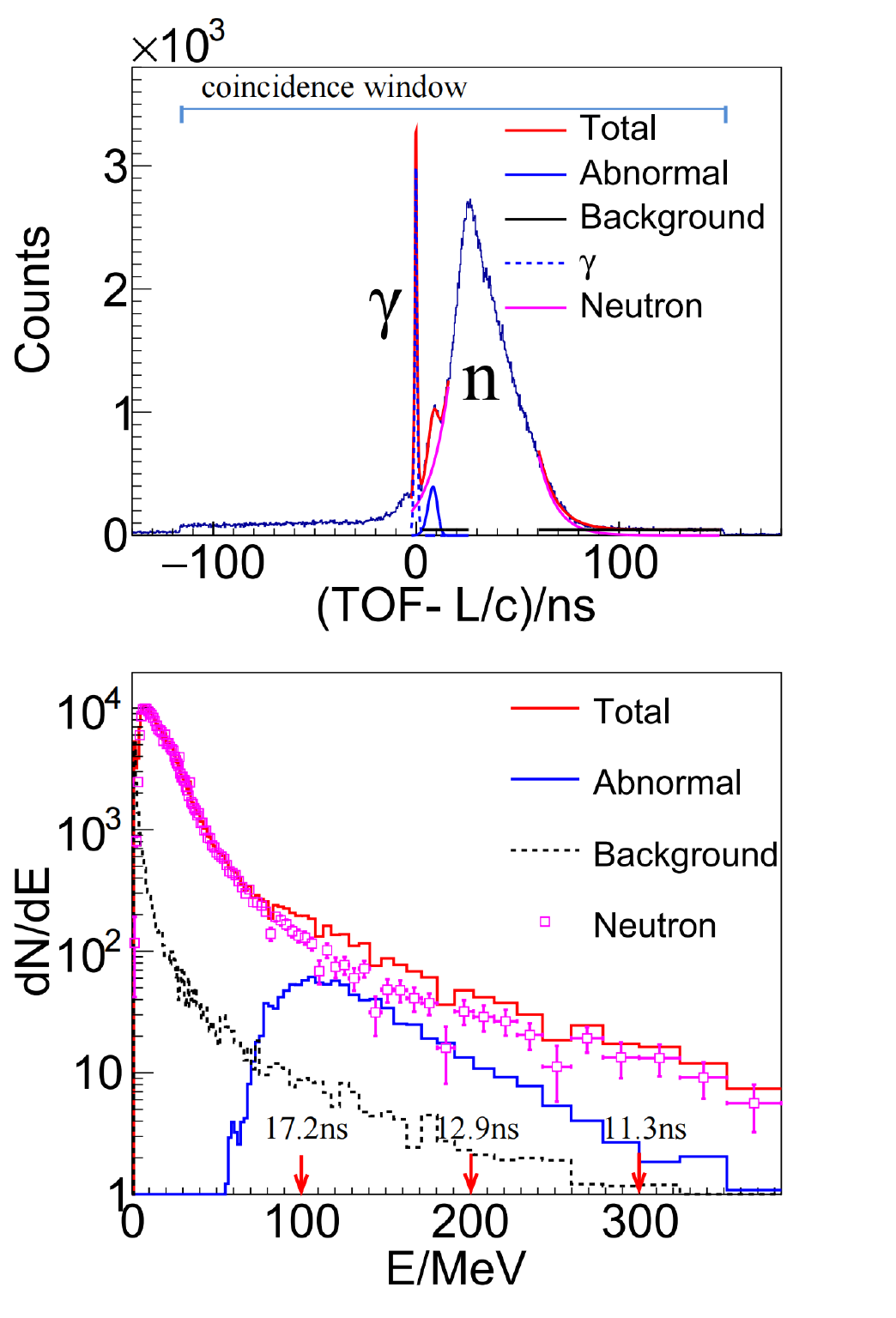}
\caption{(Color online) (a) the TOF  distribution for the $\rm 5^{th}$ unit, the TOF of $\gamma$ rays $L/c$  is subtracted on the abscissa. (b) the energy spectrum derived from the TOF spectrum for the  same unit.}
\label{fig10}
\end{figure}

\begin{figure}[!htb]
\includegraphics[width=0.85\hsize]{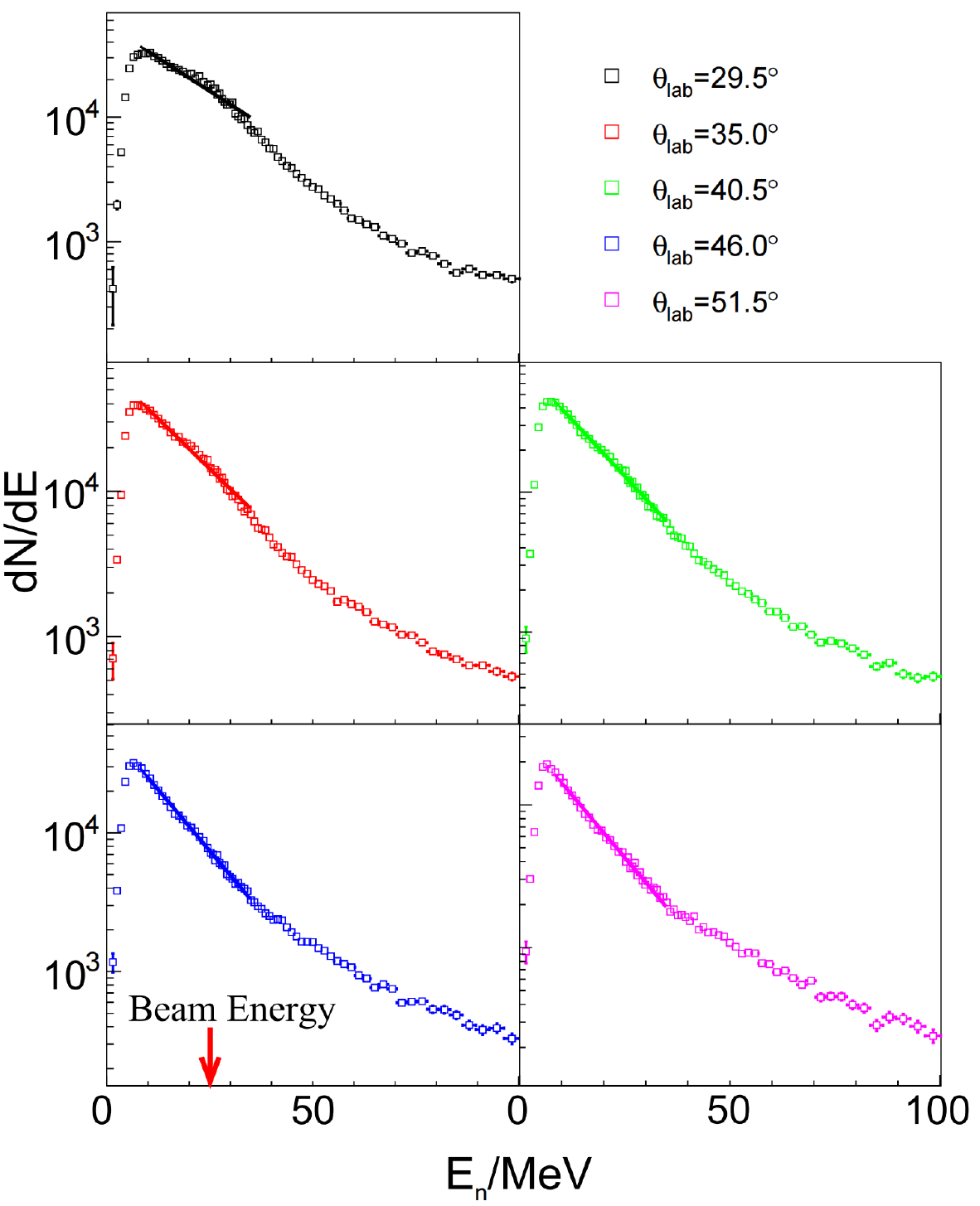}
\caption{The neutron energy spectra at different $\theta_{\rm lab}$.}
\label{fig10}
\end{figure}

Figure 11 presents the neutron spectra of the five columns without correcting the detection efficiency. The  abnormal peak and the background contributions are subtracted. The $\theta_{\rm lab}$ of each spectrum is averaged over all the units in the column. It is shown that the spectra exhibit approximately the exponential descending tail.  An exponential function $\exp(-E_{\rm n}/T)$ was used to fit the energy spectra from 8 to 35 MeV. With the increase of $\theta_{\rm lab}$, the slope parameter $T$  decreases with $\theta_{\rm lab}$, listed  here (20.40$_{-0.07}^{+0.07}$ MeV,15.77$_{-0.03}^{+0.05}$ MeV, 13.48$_{-0.03}^{+0.03}$ MeV, 12.33$_{-0.04}^{+0.03}$ MeV, 12.44$_{-0.05}^{+0.05}$ MeV) for the 5 angles, respectively. Meanwhile, the yield of the neutrons also decreases slightly with $\theta_{\rm lab}$. Both trends are consistent with the kinetic feature of heavy ion collision in the energy domain. In addition, as shown in Fig. 11, a large number of neutrons are beyond the beam energy. It indicates that the high momentum tail of nucleons may exist as a signature of short range correlation in nuclei. Furthermore, a high statistics of two-body events  of charged particles and neutrons can be utilized to analyze  the neutron-neutron and neutron-proton correlation functions.  Given the well detected neutrons shown here, along with the detection ability of the charged particles demonstrated in \cite{CSHINE1,SSD}, the feasibility of the aforementioned physics goals are promising and  further prospective investigations are required. 



\section{Summary}  \label{sec. IV}
A neutron array consisting of 20 plastic scintillator units has been assembled and mounted at the compact spectrometer of heavy ion experiments (CSHINE) to measure the fast neutron from heavy ion reactions at Fermi energies. In this work, the design, simulation, off-line and in-beam test results are reported. The inherent time resolution of 212 ps is obtained by cosmic ray test, and the  n-$\gamma$ discrimination by the time of flight information is verified by radioactive source and in-beam test. The efficiency curve has been calculated using the Geant 4 simulation tool packages. The array has been operated in the  experiment of $^{124}$Sn+$^{124}$Sn at beam energy of $E_{\rm beam}=25$ MeV/u, and the neutron energy spectra have been obtained in the angular range of $30^\circ \le \theta_{\rm lab} \le 50^\circ$. A large number of the events with the neutron energy $E_{\rm n}>E_{\rm beam}$  has been observed. Various physical goals, including the nucleon-nucleon correlation functions and the high momentum of the nucleons in nuclei, are feasible for further studies.


\end{document}